\documentclass[12pt]{iopart}

\usepackage{subfigure}
\usepackage{epsfig}

\begin{document}

\title[How well does pQCD describe strangeness in $p+p$ ?]{How well does NLO pQCD describe strangeness
production in $p+p$ collisions at $\sqrt{s}$= 200 GeV in STAR?}

\author{Mark Heinz  \footnote[1]{Electronic address (mheinz@bnl.gov)}
for the STAR Collaboration\footnote[2]{For the full author list and
acknowledgements, see Appendix ``Collaborations" in this volume.}}

\address{Laboratory of High Energy Physics, University of Bern,
CH-3012 Switzerland}

\begin{abstract}
We present measurements of the transverse momentum spectra for
$\mathrm{K^{0}_{S}}$, $\Lambda$, $\Xi$ and their antiparticles in
p+p collisions at $\sqrt{s}=200 GeV$. The extracted mid-rapidity
yields and $\langle \mathrm{p_{T}}\rangle$ are in agreement with
previous $p+\bar{p}$ experiments while they have smaller statistical
errors. We compare the measured spectra for $\mathrm{K^{0}_{S}}$ and
$\Lambda$ to the latest available calculations from NLO pQCD and see
good agreement for the $\mathrm{K^{0}_{S}}$ above 1.5 GeV/c.

\end{abstract}

\section{Introduction}
We report results from the 2002 p+p running with the STAR experiment
at RHIC. We present the high statistics measurement of
$\mathrm{K^{0}_{S}}$, $\Lambda$ and $\Xi$ particles and obtain the
yield and $\langle \mathrm{p_{T}} \rangle$ for each species. A brief
discussion of appropriate functions that were used to parameterize
the particle spectra in order to extrapolate the measurement at low
transverse momentum will ensue.
\\
Previously we have compared our measurement to the PYTHIA model
incorporating leading-order (LO) pQCD processes and have noted large
discrepancies between our data and the model \cite{MH2005}. In this
paper we compare to next-to-leading order (NLO) calculations which
exhibit a better agreement with our $\mathrm{K^{0}_{S}}$ data, but
still fail to describe the $\Lambda$.
\\
Furthermore, we study the dependency of $\langle \mathrm{p_{T}}
\rangle$ as a function of charged particle event multiplicity
($N_{ch}$) for different particle species. It has been shown that
the high transverse momentum final state particle is mostly governed
by hard partonic processes and subsequent string fragmentation
\cite{Sjo87}. We observe a strong dependence of this part of the
spectra with respect to event multiplicity.

\section{Analysis}
The present data were reconstructed using the STAR detector system
which is described in more detail elsewhere \cite{STAR1,STAR2}. The
main detector used in this analysis is the Time Projection Chamber
(TPC) covering the full acceptance in azimuth and a large
pseudo-rapidity coverage ($\mid \eta \mid < 1.5$). A total of 14
million non-singly diffractive (NSD) events were triggered with the
STAR beam-beam counters (BBC) requiring two coincident charged
tracks at forward rapidity. Due to the particulary low track
multiplicity environment in p+p collisions only 76\% of primary
vertices are found correctly; from the remainder, 14\% are lost and
10\% are incorrectly reconstructed as demonstrated by a MC-study. Of
all triggered events, 11 million events passed the selection
criteria requiring a valid primary vertex within 100cm along the
beam-line from the center of the TPC detector. The strange particles
were identified from their weak decay to charged daughter particles.
The following decay channels and the corresponding anti-particles
were analyzed: $\mathrm{K^{0}_{S}} \rightarrow \pi^{+} + \pi^{-}$
(b.r. 68.6\%), $\Lambda \rightarrow p + \pi^{-}$(b.r. 63.9\%)
,$\Xi^{-} \rightarrow \Lambda + \pi^{-}$(b.r. 99.9\%). Particle
identification of daughters was achieved by requiring the dE/dx to
fall within the 3$\sigma$-bands of theoretical Bethe-Bloch
parameterizations. Further background in the invariant mass was
removed by applying topological cuts to the decay geometry.
Corrections for acceptance and particle reconstruction efficiency
were obtained by a Monte-Carlo based method of embedding simulated
particle decays into real events and comparing the number of
simulated and reconstructed particles in each $p_{T}$-bin.

\section{Results and Discussion}

\begin{figure}[h]
\begin{center}
\epsfig{figure=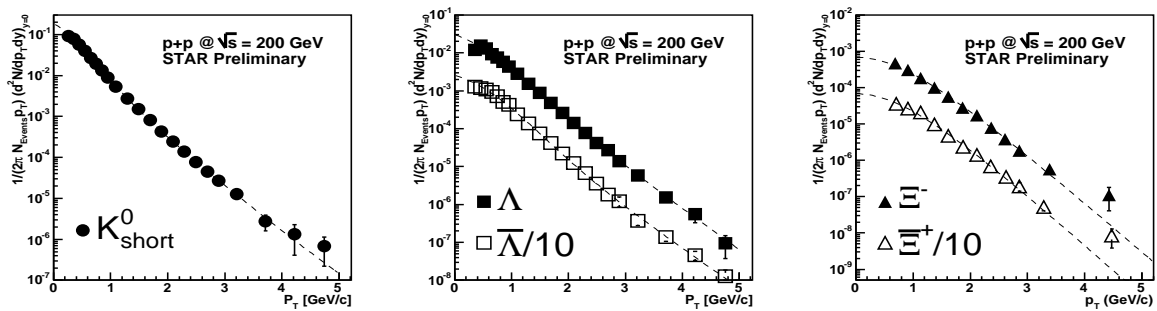, height=4.5cm, width=16cm}
\caption{Minimum-bias, non-feeddown corrected spectra for
$\mathrm{K^{0}_{S}}$ (left), $\Lambda$ (center) and $\Xi$ (right)
with ($\mid{y}\mid < 0.5$) from $p+p$ at $\sqrt{s} = 200 $GeV. The
errors in the plot are statistical only. The fits are described in
the text.} \label{fig:AllSpec}
\end{center}
\end{figure}

In Figure \ref{fig:AllSpec} corrected inclusive spectra are shown
for $\mathrm{K^{0}_{S}}$, $\Lambda$, $\Xi$ and their respective
antiparticles. The particle acceptance at mid-rapidity
($\mid$y$\mid$ $\leq$ 0.5) in the TPC starts at a transverse
momentum of 0.2 GeV/c for $\mathrm{K^{0}_{S}}$, 0.3 GeV/c for the
$\Lambda$ and 0.5 GeV/c for the $\Xi$. In order to extract the
$\langle \mathrm{p_{T}} \rangle$ and yield at mid-rapidity, a
parameterization to the spectra has to be applied to extrapolate the
measurement to cover the full $\mathrm{p_{T}}$-range. In contrast to
previous $p+\bar{p}$ experiments \cite{UA1}, which used either a
single exponential function in transverse mass or power-law
functions, we found that a combination of these functions is more
effective in fitting the singly-strange particles. Composite fits,
using an exponential function in $m_{T}$ at low $\mathrm{p_{T}}$ and
power-law functions at high $\mathrm{p_{T}}$ yielded the lowest
$\chi^{2}$ and were used to extract the values for yield and
$\langle \mathrm{p_{T}} \rangle$ of $\mathrm{K^{0}_{S}}$ and
$\Lambda$. For the $\Xi$ spectrum the limited coverage at low
$\mathrm{p_{T}}$ makes it insensitive to the different functions and
thus only an exponential function in $m_{T}$ was used. The values as
shown in table \ref{tab:ResultsUA5} are in agreement with the
measurements by UA5 when scaled using a rapidity distribution
obtained from simulation.

\begin{table}[h]
\begin{center}
\begin{tabular}{|c|c|c|c|c|c|}
\hline Particle  &STAR dN/dy         & UA5 dN/dy & UA5 dN/dy           &STAR $\langle \mathrm{p_{T}} \rangle$&UA5 $\langle \mathrm{p_{T}} \rangle$ \\
                 &$\mid$y$\mid <$ 0.5&           &$\mid$y$\mid <$ 0.5 &  [GeV/c]                            &  [GeV/c]\\
\hline
$\mathrm{K^{0}_{S}}$      & 0.128 $\pm$0.08    & 0.72 $\pm$0.12 \cite{UA5_87}& 0.15 $\pm$0.03  &0.603 $\pm$0.006 & 0.53+0.08,-0.06\\
\hline
$\Lambda + \bar{\Lambda}$ & 0.066 $\pm$0.006   & 0.27 $\pm$0.07 \cite{UA5_89}& 0.08 $\pm$0.02  &0.76 $\pm$0.02   & 0.8 +0.2,-0.14 \\
\hline
$\Xi + \bar{\Xi}$         & 0.0036 $\pm$0.0012 & 0.03+0.04,-0.02\cite{UA5_89} & 0.007 $\pm$0.010 &0.96 $\pm$0.05 & 0.8 +0.4,-0.2 \\
\hline
\end{tabular}
\caption {A comparison of mid-rapidity yields and $\langle
\mathrm{p_{T}} \rangle$ for $\mathrm{K^{0}_{S}}$, $\Lambda$
(feed-down corrected) and $\Xi$ and measured by STAR and UA5. UA5
measurements were made over a large rapidity interval and were
scaled down using the following factors obtained from simulation:
$\mathrm{K^{0}_{S}}$= 4.67($\mid$y$\mid <$ 3.5), $\Lambda$=3.27
($\mid$y$\mid <$ 2.0), $\Xi$=4.03 ($\mid$y$\mid <$ 3.0)}
\label{tab:ResultsUA5}
\end{center}
\end{table}

The systematical errors are mainly due to the different fit
parameterizations and amount to 8\% for $\langle \mathrm{p_{T}}
\rangle$ and 10\% on the yield for V0s and 10\% respectively 20\%
for Xis. The $\bar{\Lambda}$/$\Lambda$ ratio is 0.88$\pm$0.09 and
$\bar{\Xi}$/$\Xi$ ratio is 0.90$\pm$0.09 and both are independent of
$\mathrm{p_{T}}$. The data sample was split into event classes with
increasing mean charged particle multiplicity per unit $\eta$. For
$\mathrm{K^{0}_{S}}$ and $\Lambda$ six event classes were possible
whereas for $\Xi$ only 3 event classes could be obtained with
sufficient statistics. The goal is to study the large momentum
transfer region of the parton-parton collisions by measuring the
spectra in high multiplicity events, where this type of interaction
is expected to be more probable.

\begin{figure}[h]
\begin{center}
\epsfig{figure=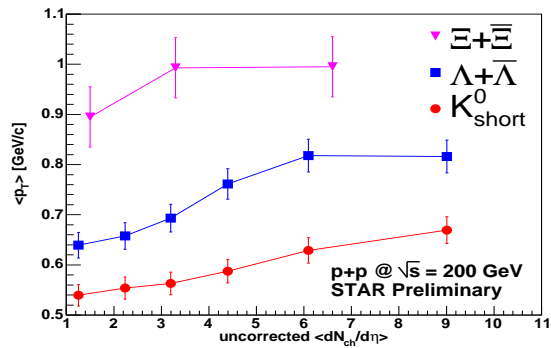, height=5cm,
width=8cm} \caption{$\langle \mathrm{p_{T}} \rangle$ vs. uncorrected
$\langle dN_{ch}/d\eta \rangle$ for $\mathrm{K^{0}_{S}}$ (circles),
$\Lambda$ (squares) and $\Xi$ (triangles). The errors shown are both
statistical and systematic.} \label{fig:MeanPtMult}
\end{center}
\end{figure}

Figure \ref{fig:MeanPtMult} presents the $\langle \mathrm{p_{T}}
\rangle$ vs $\langle dN_{ch}/d\eta \rangle$ for
$\mathrm{K^{0}_{S}}$, $\Lambda$, and $\Xi$ measured by STAR. A rise
in $\langle \mathrm{p_{T}} \rangle$ with increasing $N_{ch}$ is
observed and the trend is stronger for the $\Lambda$ than for the
$\mathrm{K^{0}_{S}}$. Unfortunately the large syst. errors on the
$\Xi$ $\langle \mathrm{p_{T}} \rangle$ measurement do not allow any
conclusions at this point. Several authors have attributed this
phenomenon to the increased number of large momentum transfer
parton-parton collisions that produce mini-jets in the high
multiplicity events \cite{Gyu92,UA1}.

\section{Comparison to NLO model calculations}

\begin{figure}[h]
\begin{center}
\epsfig{figure=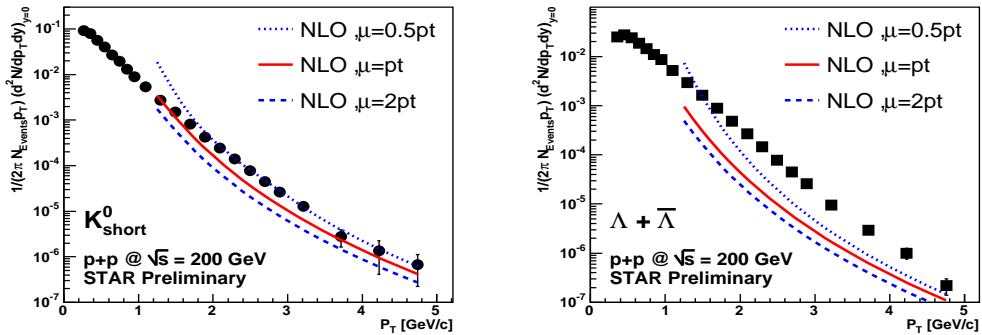, height=5cm, width=14cm}
\caption{$\mathrm{K^{0}_{S}}$ (left) and $\Lambda$ (right) particle
spectra compared to NLO pQCD calculations provided by W.Vogelsang.}
\label{fig:SpectraNLO}
\end{center}
\end{figure}

As mentioned in the introduction, we have previously attempted to
compare our data to LO pQCD models and have seen large
discrepancies. It was shown at this conference that the STAR charged
particle spectrum is well described by NLO pQCD calculations
\cite{MVL2005}. Therefore, we have obtained recent NLO pQCD
calculations for the strange particles and compared to the data in
figure \ref{fig:SpectraNLO}. These calculations use KKP
fragmentation functions for $\mathrm{K^{0}_{S}}$ \cite{KKP} and
fragmentation functions by Vogelsang et al. for $\Lambda$ \cite{WV}.
The disagreement between the calculations and the data are small for
$\mathrm{K^{0}_{S}}$ but considerably larger for the $\Lambda$. This
may be due to the higher mass of the $\Lambda$, where the massless
quark formalism breaks down and the (m/$\mathrm{p_{T}}$)-scale
approximations become non-negligible. Bourrely and Soffer have
calculated alternative fragmentation functions for octet baryons
which need to be tested \cite{Bourrely}.

\section{Summary}
The STAR experiment has made the first high statistics measurement
of mid-rapidity $\mathrm{K^{0}_{S}}$, $\Lambda$ and $\bar{\Lambda}$,
$\Xi$ and $\bar{\Xi}$ in $p+p$ collisions at $\sqrt{s}$ = 200 GeV.
The results agree with those made by the UA5 collaboration for
$p+\bar{p}$ collisions at the same energy. The ratio of
$\bar{\Lambda}$/$\Lambda$ and $\bar{\Xi}$/$\Xi$ suggests a small net
baryon number at mid-rapidity.
\\
Furthermore, we show that NLO pQCD calculations reproduce the
experimental data much better than LO-calculations. The model agrees
reasonably well for $\mathrm{K^{0}_{S}}$ above 1.5 GeV/c but still
fails to reproduce the shape of the $\Lambda$ spectra.
\\
Finally, we have undertaken studies to understand the change in
$\langle \mathrm{p_{T}} \rangle$ of different strange particle
species with increasing event multiplicity in an attempt to
understand the flavor dependance of fragmenting mini-jets in high
multiplicity event samples.

\end{document}